\documentclass[conference]{IEEEtran}
\usepackage[english]{babel}
\usepackage{fancyhdr}
\usepackage{amsmath}
\usepackage{amssymb}
\usepackage{graphicx}
\usepackage{adjustbox}
\usepackage{algorithm}
\usepackage{algorithmic}
\usepackage{mathtools}
\newcommand\Mark[1]{\textsuperscript{#1}}

\ifCLASSINFOpdf
\else
\fi
\begin{document}
%
\title{B-FICA: BlockChain based Framework for \\ Auto-insurance Claim and Adjudication\\[.75ex]
   {\normalfont\large 
   Chuka Oham\Mark{1,2}, Raja Jurdak\Mark{2}, Salil S. Kanhere\Mark{1}, Ali Dorri\Mark{1,2}, Sanjay Jha\Mark{1}%
   }\\[-1.5ex]
}
\author{
    \IEEEauthorblockA{%
       \Mark{1}University of New South Wales, Sydney \Mark{2}CSIRO, data61
    }

}


%


\maketitle


\begin{abstract}
In this paper, we propose a partitioned BlockChain based Framework for Auto-insurance Claims and Adjudication (B-FICA) for CAVs that tracks both sensor data and entity interactions with two-sided verification. B-FICA uses permissioned BC with two partitions to share information on a need to know basis. It also uses multi-signed transactions for proof of execution of instructions, for reliability and auditability and also uses a dynamic lightweight consensus and validation protocol to prevent evidence alteration. Qualitative evaluation shows that B-FICA is resilient to several security attacks from potential liable entities. Finally, simulations show that compared to the state of the art, B-FICA reduces processing time and its delay overhead is negligible for practical scenarios and at marginal security cost. \\
\end{abstract}


%
\IEEEpeerreviewmaketitle

\section{Introduction}
Connected and automated vehicles (\textit{CAVs}) are instrumented with wide arrays of specialised computers called Electronic Control Units (ECUs), sensors and connecting technologies to facilitate independent driving decisions and enable better perception of the environment to avert road transportation hazards. The ability of  \textit{CAVs} to make some or all driving decisions disrupts the current auto insurance claims and adjudication process where a physical driver is solely blamed for an accident. This disruption enables shared liability among multiple entities interacting with a \textit{CAV} including the Automotive Manufacturer, Service Technician and Software Provider and therefore makes adjudication challenging. Consider the fatal \textit{CAV} accident scenario described in [1]. In this scenario, assigning blame was arduous due to lack of transparency in the adjudication process. Furthermore, with the existing claims processing model, a claimant in this scenario would have to painstakingly wait for about 4-12 months for an insurer to make a decision based on the circumstances of submitted claim [29].  \\
Given the potential to assign blame to multiple entities, the adjudication process for \textit{CAVs} presents further interesting challenges, including: (1) the motivation of potential entities to evade liability by altering evidence; (2) keeping track of relevant interactions between a \textit{CAV} and other potential liable entities; (3) providing comprehensive evidence for adjudication; and (4) ensuring data accessibility. \\
Previous works that address vehicular forensics significantly rely on data stored in the Event Data Recorder (EDR); an automotive black-box replica installed in the vehicle for recording pre and post collision data [2] [3] [4]. Several challenges however persist with \textit{CAVs}. First, EDR data does not record all the relevant interactions of \textit{CAVs} with other entities. This prevents existing approaches from making liability decisions that consider these interactions. Second, modern day vehicles with communication capabilities log EDR data in central servers [8] introducing a single point of failure. Third, an evaluation on proposed techniques for vehicular forensics has been missing. \\
In this paper, we argue that BlockChain (BC) [5] technology has the potential to address the aforementioned challenges. BC is an immutable and distributed ledger technology that provides verifiable record of transactions in the form of an interconnected series of data blocks. BC can be public or permissioned [25] to differentiate user capabilities including who has right to participate in the BC network. Public BC such as for BitCoin [5], enables every participant to create transactions (BC communications), collaboratively verify them and append validated transactions to the ledger. Permissioned BC such as Hyperledger [6] on the other hand restricts participation by inviting participants to join and allows only selected participants to maintain the shared ledger. \\
BC replaces centralization with a trustless consensus which ensures that no single entity can assume absolute control of the BC and its contents. The decentralized trust provided by BC is well-suited for the \textit{CAV} adjudication process.  Also, the distributed structure of BC delivers robustness to a single point of failure. Therefore, we propose a BC Framework for auto-Insurance Claims and Adjudication (B-FICA). B-FICA is a vehicular forensic system for \textit{CAVs} that facilitates the collection of relevant evidence from potential liable entities. It consists of two main tiers namely, operational and decision. Given known identities of interacting entities in the auto insurance adjudication model, we use a permissioned BC to ensure that interacting entities are only privy to communications (transactions) they need to know. The operational partition comprises \textit{CAVs}, insurance companies, auto manufacturers, software providers and service technicians. As evidence, the operational partition stores specific instructions (software updates, change of CAV parts) from an auto manufacturer or service technician to a CAV owner as well as data generated by a \textit{CAV} when it is in an accident or behaves in an unexpected way such as over speeding or hard slamming of brakes. The aim of tracking interactions with the \textit{CAV} is to identify cases of neglect to instructions, while the \textit{CAV} data maintains a historical record of the \textit{CAV} performance and circumstances surrounding an accident. To ensure scalability, the operational BC partition is managed by entities with significant communication and storage capabilities such as auto manufacturer, insurance company and service technicians. The decision partition comprises the legal authorities including the police and courts, the government transport authorities, insurance companies and automotive manufacturers. In the decision partition, requests for complimentary evidence such as witnesses testimonies are made by insurance companies to the legal and government transport authorities. Also, contestable decisions reached in the operational partition are resolved in the decision partition.\\
Given the infrequent rate of data generation and vested interests of potential liable entities in adjudication model, B-FICA introduces a dynamic lightweight consensus and block validation protocol and a proof of interaction-execution mechanism. The dynamic lightweight consensus and block validation protocol allows BC managers to update the BC state for every received data contributing to evidence to prevent the alteration of evidence. The proof of interaction-execution mechanism offers non-repudiation, allows BC managers confirm the status of a BC communication and helps adjudicators identify if an accident was due to negligence. \\ 
\textit{The key contributions of this paper are summarized below:}
\textbf{(1)} We present B-FICA; a partitioned BC framework based on permissioned BC technology for obtaining comprehensive evidence to facilitate adjudication and claims processing. We use realistic scenarios to highlight the efficacy of our proposal. \\
\textbf{(2)} Our BC based framework is tailored to meet the requirement for the CAV adjudication process. We incorporate a proof of interaction-execution mechanism and a dynamic lightweight consensus and block validation protocol to mitigate malicious actions by participants. \\
\textbf{(3)} We conduct a qualitative analysis of B-FICA to evaluate the resilience of B-FICA against identified attack capabilities of malicious actors. Also, we conduct a comparative evaluation of B-FICA with existing approaches. \\
\textbf{(4)} We characterize the performance of B-FICA via extensive simulations using NS3 simulator and comparatively evaluate our proposal against an existing proposal using time overhead as a performance measure. \\
This paper is an extension of our preliminary ideas presented in our previous work [7]. Here we include a use case, algorithm and evaluations. \\
The rest of the paper is organized as follows. The proposed architecture and components are described in Section 2. We present an overview of transactions in Section 3. Evaluation results are described in Section 4. Section 5 presents related works on vehicular forensics and BC applications and Section 6 concludes the paper and outlines future work. 
\section{BlockChain based Framework for Insurance Claim and Adjudication (B-FICA)}
In this section, we discuss the details of the BC partitions. We begin by defining key concepts. \\
\textbf{Transactions:} Transactions are the basic communication primitive in BC for information exchange among entities in B-FICA.\\
\textbf{Partitions:} Partitions are communication segments that allow intended communication participants to exchange relevant transactions contributing to evidence. \\
\textbf{Dynamic Block (\textit{dBlock}):} An incomplete block containing validated transactions that allows validators reach consensus on the current state of the BC. \\
\textbf{Proposers:} Proposers are entities that initiate and submit transactions to the BC. \\
\textbf{Validators:} Validators are entities that are responsible for distributedly managing the BC by processing incoming and outgoing transactions in their partitions.\\
The proposed framework is presented in Figure 1. It describes the interaction between key components in B-FICA. 
\begin{figure*}[h]
\centering
\includegraphics[width=0.8\textwidth]{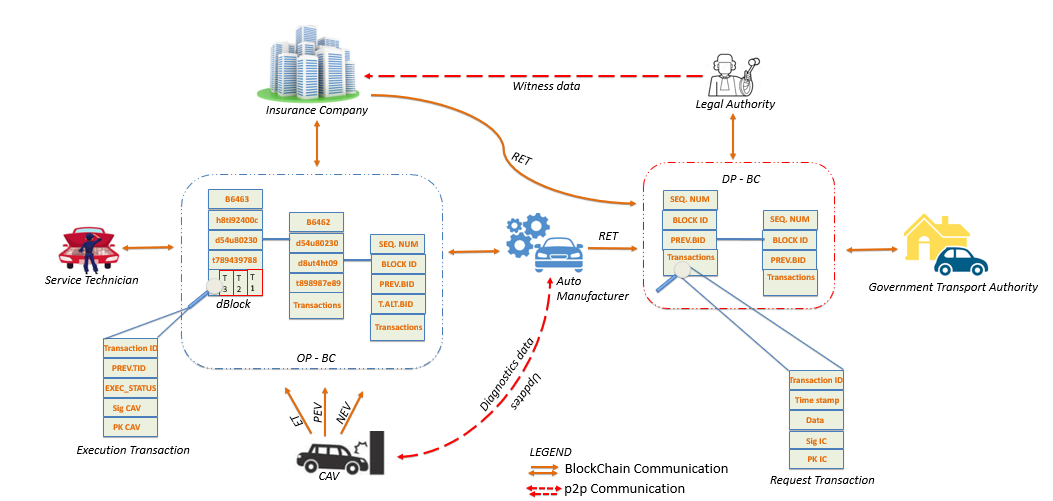}
\caption{Proposed Partitioned BC Based Framework.}
\end{figure*}
\subsection{BC Partitions}
The partitions in our framework are the operational and decision partitions. These partitions are intended to facilitate the communication of relevant information across our BC network. In the operational partition, transactions contributing to evidence are generated and stored in the dynamic block. In the decision partition, complimentary evidence is presented to insurance companies to facilitate compensation payments, contestable liability decisions are resolved and liability decisions involving multiple vehicles is made. \\
For BC communications, we utilize existing Public Key Infrastructure (PKI) such as a Certificate Authority (CA) to issue unique digital identities to BC participants in the operational partition and decision partition. The certificate or verification component of the CA is stored in the genesis block which serves as a starting point of the BC and used to authenticate and verify transactions.  \\
For communications initiated by a validator and completed by a \textit{CAV}, known keys of transaction participants are used to authenticate the transactions. However for transactions generated by a \textit{CAV} such as the accident-related data, we utilize changeable keys. This is because a \textit{CAV} is envisaged to also collect significant personal identifiable information of the owner and using known keys for communication might threaten the privacy of the vehicle owner. We utilize a pseudonym based privacy method similar to the approach in [30, 31] to address this challenge. In this approach, it is assumed that at the initial registration of a \textit{CAV} by a government transport authority, the \textit{CAV} is issued a set of pseudonyms for the purpose of communicating accident-related data. To prevent data misuse, a law enforcement agency registers for this service and obtains requisite communication credentials to participate in the process of uncovering the real identities of \textit{CAVs} for adjudication.  This approach does not require a vehicle to store significant amounts of short-lived pseudonymous certificate as akin to the public key certificate (PKC) approach. It offers user-controlled privacy to the \textit{CAV} owner by allowing a vehicle to change its pseudonyms at any given time. 
\subsubsection{Operational Partition BC (OP-BC)}
The operational partition comprises all potential liable entities including the \textit{CAV}, auto manufacturer, service technician in addition to the insurance company responsible for paying compensations. For simplicity we focus on single entities for each of these however, the framework is generalisable to the case when there are several of each entity. \\
We assume that an auto manufacturer provides the software update needs of a \textit{CAV}. We also assume that a \textit{CAV} is 5G-enabled [30] and has a tamper-resistant storage for storing accident-related data such as the precise location, speed, time of event, video and picture data of an accident including data received from witnesses ($W_{i}$) via vehicular communication.\\ 
The proposers in the operational partition include the auto manufacturer, service technician and the CAV while the validators include the auto manufacturer, service technician and the insurance company. Validators are selected based on their possession of sufficient resources for processing transactions and blocks. Also, given that validators could also send transactions in the OP-BC, we utilize a dual signature transaction mechanism to ensure that transactions initiated by a validator are acknowledged by the recipient before they are sent to the BC. This  prevents the possibility of a validator generating fake transactions to evade liability.\\
In the operational BC, transactions are either signed by a single participant called single sign transactions or by multiple participants called multiSig transactions. The key transactions in OP-BC allow participants to report safety events, provide primary evidence, provide specific instructions, or notify execution of instructions. Section III provides a detailed description of these transactions.  
Transactions in the operational partition are broadcast to the BC network and stored in the OP-BC. Each block in the OP-BC consists of two parts namely the block header and transactions. A block header contains the following: block sequence number (\textit{SEQ. NUM}), block ID (\textit{BLOCK ID}) which is the hash of the block, the hash of the previous block (\textit{PREV.BID}) which ensures immutability and the hash of the transaction altering the dynamic block (\textit{T.ALT.BID}) which is a pointer to the last transaction that changes the state of \textit{dblock}. For \textit{dblock}, the \textit{blockID} changes for every successfully validated transaction until dBlock reaches maximum block allowance ($B_{Max}$) and appended to OP-BC. The dynamic \textit{blockID} is also used to ensure immutability in OP-BC. The motivation for a dynamic block in our framework is primarily due to the vested interest of potential liable entities in the outcome of an adjudication decision and secondarily due to the infrequent rate of data generation for adjudication. The waiting for transactions in the running pool to reach $B_{Max}$ before validation offers rogue OP-BC validators’ ample time to alter transactions to evade liability. For example, a rogue auto manufacturer who observes that a product defect was responsible for the accident could attempt to alter transactions in the running pool to evade liability. However, with \textit{dblock}, the updated \textit{blockID} would detect this attack. \\  
\textbf{Transaction verification and dynamic block validation:} Once a transaction is sent to the OP-BC, the process of verification and validation begins by a validator checking first that the transaction generator is an authorized OP-BC participant. Next it checks to see if the transaction is complete. For example if a multiSig transaction has the signatures of all transaction participants and finally if the transaction is not a duplicate transaction sent by a particular participant. Algorithm 1 outlines the transaction verification and validation process in OP-BC. 
\begin{algorithm}[h]
\hspace*{\algorithmicindent} \textbf{Input: Transaction} \\
\hspace*{\algorithmicindent} \textbf{Output: new \textit{dBlock ID}} 
\begin{algorithmic} [1]
\STATE \textit{Verify transaction}:
\STATE  Verify  $PK_{AV}$ ;
\IF {$PK_{AV}$ is not valid for \textit{OP-BC}}
\STATE $Transaction$ is invalid ;
\STATE \textbf{reject Transaction} ;
\ELSE
\STATE \textit{Validate transaction}:
\STATE \textit{Compute new cBlock ID (ndB)} ;
\STATE \textit{ndB} = $Curr.T_{ID}$ + $dBlock_{ID}$ ;
\STATE Verify \textit{Computational consistency} ;
\STATE For every successfully verified \textbf{Transaction} ;
\WHILE {$dBlock < B_{Max}$}
\STATE \textit{ndB} = $Curr.T_{ID}$ + $dBlock_{ID}$ ;
\ENDWHILE
\ENDIF
\end{algorithmic}
\caption{Transaction Verification and Dynamic Validation }
\end{algorithm}
In the OP-BC, we assume that validators use predefined known keys to generate a dynamic block (\textit{dBlock}). Upon a successfully verified transaction, the transaction is validated in \textit{dBlock} by computing a new \textit{dBlock ID} (lines 9, 10). This is done by computing the hash of the successfully verified transaction \textit{(Curr.Tid)} and the hash of \textit{dBlock}. Once computed, validators compare their values to ensure computational consistency and reach consensus on the state of \textit{dBlock}. The validation process continues until \textit{dBlock} reaches the maximum block allowance ($B_{Max}$). Once reached, \textit{dBlock} is appended in the BC. In the case where different \textit{dBlock} output is derived, validators disregard current computation and revert to the last computation where \textit{dBlock} \textit{ID} was consistent across all validators. Then using the last validated transaction identified by the \textit{T.ALT.BID} identifier, OP-BC validators re-initiate the verification and validation process up until the current transaction to be validated by checking transaction timestamps. Repeating this process using the last successfully validated transaction will allow validators to reach consensus on the state of OP-BC and detect the likelihood of an attack. If however wrong computation persists leading to contestable decisions, \textit{dBlock} is presented to the decision partition validators for resolution. 
\subsubsection{Decision Partition BC (DP-BC)}
The decision partition BC integrates auto manufacturers, insurance companies, legal authorities and government transport authorities. The proposers in the decision partition include insurance companies and auto manufacturers. On the one hand, the insurance companies initiate a request for complimentary evidence to facilitate processing of insurance claims. On the other hand, auto manufacturers also initiate a request to identify a liable \textit{CAV} in a multi-collision scenario. The validators in DP-BC are the government transport authority and the legal authorities who collaborate to reveal to identity of a liable \textit{CAV} in a collision scenario. The legal authority also provides complimentary evidence to an insurance company to facilitate liability attribution and claims processing. Each block in the DP-BC consists of a block header and transaction as with the OP-BC. However, the block header in DP-BC as opposed to the OP-BC does not include the hash of the transaction altering the block. This is because validation of transaction in the DP-BC occurs only when the running pool of transactions have reached the maximum block allowance ($B_{Max}$). Also, vested interest in the outcome of liability decisions is removed in the DP-BC as DP-BC validators are not considered liable entities in B-FICA. Once transactions in the running pool of a validator have reached $B_{Max}$, the validator creates a new block and the new block is appended to the DP-BC when consensus is reached on the state of the new block. \\
The request for evidence is the sole transaction stored in the DP-BC. The response to the request is sent as a unicast to the request initiator and not stored in the DP-BC. This is to ensure that sensitive privacy related information is not made public knowledge. \\ 
\textbf{Transaction verification and validation:} Upon the receipt of a \textit{RET} transaction, validators verify the authenticity of the transaction by verifying the signatures of proposers. Next, they verify that transaction is complete. Validators repeat this process for subsequent transactions until their running pool reaches maximum block allowance ($B_{Max}$). Next, validators create a block containing $B_{Max}$ transactions  and repeat the verification process to ensure that transactions are unique and verifiable. Once successful, the validation process begins by computing the hash of all transactions in the current block. Computed values are compared for consistency and the successfully validated block is appended to DP-BC.
\section{Overview of Transactions}
Having discussed the details in the BC-partitions, we discuss the details of interactions in each partition facilitated by the different kind of transactions. Transactions generated by proposers in our framework are secured using cryptographic hash functions (SHA256), digital signatures and asymmetric encryption.
\subsection{Operational BC Transactions}
Communication between entities in the OP-BC partition are referred to as operational transactions (\textit{OP-T}). \textit{OP-T} are evidence used for making liability decisions. They include relevant interactions between liable entities as well as collision data to identify cases of product, service and negligence liability [9]. Product liability refers to scenarios where blame is assigned to an auto manufacturer for product defect. Service liability refers to scenarios where the identified last action of a service technician caused the accident. Negligence liability refers to scenarios where failure by a vehicle owner to adhere to instructions from an auto manufacturer or service technician is found to be responsible for the accident. To capture these liability scenarios, the following transactions have been defined for the operational BC partition. \\
\textbf{Event Safety Evidence (ESE):} A \textit{CAV} sends this transaction to the OP-BC when an unexpected vehicle behaviour occurs such as hard brake slam, detection of wrong way driving and the detection of slippery road condition [10]. ESE is predicated on safety messages exchanged in vehicular networks [10]. ESE is a single sign transaction. Insurance companies use ESE to have a historical perception of the behaviour of a \textit{CAV}.\\
\textbf{Primary Evidence Transaction (PET):} A \textit{CAV} initiates the single sign PET transaction when an accident occurs. PET contains requisite data describing the accident. When an accident occurs, the \textit{CAV}CAV records all the required data, defined by [11], for making adjudication decisions. Details of \textit{PET} is described in Section III-C.\\ 
\textbf{Notification Evidence Transaction (NET):} \textit{NET} is initiated by a validator requesting the execution of specific action by a \textit{CAV} such as carrying out a software update. \textit{NET} is a multiSigned transaction requiring the signatures of the instruction issuer and the \textit{CAV} to be complete. The multiple signatures for \textit{NET} help prevent false notification claims by malicious validators when an accident occurs. For a software update instruction, the transaction includes the hash of the transaction, hash of the update file, dual signatures and a metadata field which gives details of the update file and a pointer to the location of the file.\\
\textbf{Execution Transaction (ET):} \textit{ET} is initiated by a \textit{CAV} as a single sign transaction in response to a \textit{NET} from an instruction issuer. \textit{ET} is linked directly to the \textit{NET} sent by a specific instruction issuer. The content of \textit{ET} includes the hash of \textit{ET}, the hash of the \textit{NET} it is responding to and an execution status field which is a binary indicator of execution success or failure. If however, a \textit{CAV} owner fails to execute the transaction and claims otherwise, the current firmware version of the updated device in the \textit{CAV} could be retrieved for adjudication purpose. We assume that the retrieved file shows the time stamp of the execution which is the time the update was installed by \textit{CAV} owner. Upon retrieval, the hash of the file is computed and compared with the hash of the update file in the BC. If computation differs, \textit{CAV} owner is blamed for negligence.                                                                                                                                
\subsection{Decision Partition BC Transaction}
In the DP-BC partition, communications are directed towards providing more evidence to an insurance company to facilitate compensation payments and identify a liable vehicle in a multi vehicle collision scenario. To make these decisions, both insurance company and auto manufacturer initiate a request transaction. \\
\textbf{Request Transaction:} In a given collision scenario involving a single or multiple vehicles, RET is sent by an insurance company after receiving the primary evidence transaction from a \textit{CAV}. \textit{RET} includes the hash of the \textit{RET}, collision data in the primary evidence and the signature of the insurance company. In response to this request the DP-BC validators provide complimentary evidence as unicast to the insurance company. 
\subsection{Usage Scenario}
We consider a rear-end collision event. Adjudication decisions in the state of the art dispute settlement model in this scenario attributes blame vehicles behind the leading vehicle irrespective of the circumstance. This possibility motivates a rogue leading man-machine driven \textit{CAV} to  intentionally cause a rear-end collision to receive compensations [12, 13]. We assume a three vehicles ($CAV_{1}$, $CAV_{2}$ and $CAV_{3}$) collision scenario and classify vehicles involved in the collision as host and witnesses. For example, $CAV_{1}$ is a host vehicle when it submits \textit{PET} to OP-BC and becomes a witness when its data is provided as complimentary evidence or used to make liability decisions in multi vehicle collision scenarios. Decision making occurs at two levels. First in DP-BC and next in the OP-BC. In the first level, liability is attributed to a liable \textit{CAV} by DP-BC validators. Once a liable vehicle is identified, the second level decision is reached in OP-BC using operational transactions and complimentary evidence obtained by the insurance company. \\
In this scenario, the process of making the 2-level liability decisions follows three distinct phases namely evidence generation, validation and adjudication. In the evidence generation phase, \textit{CAVs} generate \textit{PET} by collating accident data. In the evidence validation phase, the validators in the operational partition confirms that \textit{CAVs} are valid BC participants and validates \textit{PET}. After this, OP-BC validators issue a request transaction (\textit{RET}) to DP-BC validators for first level decision making. In the adjudication phase, DP-BC validators reach decision on the liable \textit{CAV} and provide complementary evidence to the insurance company to facilitate second level decisions. \\
\textbf{Evidence generation:} After the collision, $CAV_{1}$, $CAV_{2}$ and $CAV_{3}$ independently record their perception of the collision event and each CAV receives an encrypted version of other CAVs perception. Once received, the vehicles generates \textit{PET} and sends it to the OP-BC. For $CAV_{1}$, \textit{PET} details include:
\begingroup
    \fontsize{8pt}{12pt}
    \begin{align*}  
        PET =[T_{id} || T_{data} || PK_{CAV_1} || Sig_{CAV_1}]
    \end{align*}
\endgroup
$T_{id}$ is the hash of the transaction, $T_{data}$ is the transaction data which includes accident related data in $CAV_{1}$ tamper storage device. $PK_{CAV_1}$ is the public key of $CAV_{1}$ and $SIG_{CAV_1}$ is its signature on the transaction. The details of the transaction data is given below. 
\begingroup
    \fontsize{8pt}{12pt}
    \begin{align*}  
        T_{data} = [(loc, TS, VE_{px},  TS_{data},  E(CAV_{2}), E(CAV_{3}),  h(T_{data})]
    \end{align*}
\endgroup
\textit{loc} represents the location of event. \textit{TS} is the timestamp on the data which represents time of occurence. $VE_{px}$ is $CAV_{1}$ record of the event, $TS_{data}$ refers to the hash of the last recorded picture and video file contained in the tamper-proof storage device before and during the accident. We store only the hash of the video and picture files in the BC for scalability reasons because \textit{CAV} sensors such as LIDAR and cameras produce high precision memory intensive data [28]. Therefore, we assume that actual video and picture file are separately stored in a cloud storage by the \textit{CAV} owner. \textit{CAV} owner creates a public / private key pair for the cloud storage and provides public key to operational BC validators to access the file for adjudication purposes. $E(CAV_{2})$ and $E(CAV_{3})$ are the encrypted accounts of witnesses and $h(T_{data})$ is the hash of the collision data. It is used to assure its integrity. 

\textbf{Evidence Validation:} After receiving the transaction OP-BC validators execute verification checks described in Section 2 and upon a successful verification the transaction is validated in the dynamic block as described in Algorithm 1. Next, the insurance companies and auto manufacturers of $CAV_{1}$, $CAV_{2}$ and $CAV_{3}$ initiate \textit{RET} in the (DP-BC). 

\textbf{Adjudication phase:} Upon receipt of the request from proposers in the DP-BC partition, validators verify the signatures of the proposers using their public keys. Next, they perform integrity checks before analysing transaction data content to make the first level decision. Integrity check is performed by verifying that transaction data has not been altered by computing the hash of transaction data and comparing it against $h(T_{data})$ in the data content sent by proposers. If computation differs, data tampering is detected. In-addition, they decrypt encrypted data to compare the time stamp and location data across the request transactions received to ensure data consistency in both time and space. Once integrity checks are successful, DP-BC validators make first level liability decision by analysing \textit{RET} data. Once decided, complimentary evidence is presented to insurance company for liability attribution in OP-BC. 
\section{Evaluation and Discussion}
In this section, we provide qualitative and quantitative evaluations of B-FICA. First we demonstrate how B-FICA is resilient to identified attacks. Next, we comparatively evaluate B-FICA against existing proposals. Finally, we characterize the performance of B-FICA. 
\subsection{Security analysis}
It is assumed that adversaries are rational internal attackers [14] motivated to evade liability costs. More precisely, an adversary could be an auto manufacturer or service technician who could also suffer reputational cost [15] which poses a significant threat to their revenue and business continuity if held responsible for an accident. A \textit{CAV} owner could also be motivated to commit insurance fraud by staging an accident to receive compensation for injuries and damages. Table 1 summarizes the mechanisms that enable B-FICA to meet critical design requirements. Next, we discuss the adversary model and B-FICA defense mechanisms. 
\begin{table*}[ht]
\caption {Requirement for blockchain based architecture} \label{tab:title} 
\centering
\begin{tabular}{|p{1.7cm}|p{15cm}|}
\hline
Requirement & Approach \\
\hline
Authorization & Partitions restrict communication to only authorized partition participants and allows communication to occur only on a need-to-know basis.  The genesis block in each partition contains the verification credentials of a Certification Authority (CA). The verification credential is unique for both partitions thus allowing only members in a given partition to authenticate and verify transactions.  \\
\hline
 Integrity & Transaction identifier ($T_{id}$) represents the hash of a transaction and used to ensure that a transaction originating from an BC participant is not modified. Also, we hash data contents in a transaction such as the primary evidence transaction to identify cases of evidence alteration. \\
\hline
Secure storage & Transactions that have been successfully verified are committed in the BC for adjudication. Also, by validating transactions in \textit{dblock}, evidence alteration is prevented. Furthermore, by storing data on the BC and dynamically validating transactions, we prevent the possibility of a rogue validator to make evidence unavailable. \\
\hline
Non-repudiation &  Transactions in a dynamic \textit{dBlock} are signed by all transaction initiators and verified by partition validators to prevent denial of action and to ensure auditability.\\
\hline
Decentralization & No single validator can independently make decisions on what data constitutes evidence. Data contributing to evidence are collaboratively verified by validators and only successfully verified transactions can be used to make liability decisions. \\
\hline
\end{tabular}
\end{table*}
\subsubsection{Adversarial Model}
The adversarial model details the malicious capabilities of liable entities and demonstrates the resilience of B-FICA. \\
\textbf{Transaction deletion:} In the operational partition, a rogue validator could leverage the infrequent rate of transaction generation and the significant time it takes for transactions in the running pool to reach $B_{Max}$ in the OP-BC to delete implicating transactions so as to evade liability. This capability distorts the block validation process and makes adjudication challenging. This process is addressed in our framework via the dynamic block validation process where transactions are validated as they are received. The dynamic block validation results in an updated \textit{block ID} for successfully validated transaction. \\
\textbf{Sign fake transaction:} A rogue OP-BC validator could compromise a \textit{CAV}, sign a fake transaction so that the adjudication process would blame the \textit{CAV} owner for neglecting an instruction that could have prevented the accident. This attack is addressed in our framework using the transaction alter field in the dynamic block which allows all validators to keep track of the last validated transaction in a dynamic block. Also, the dynamic validation process described in Section 2 would enable validators to detect this attack. \\
\textbf{Collusion:} We define two cases of collusion attack across the BC-partitions below. \\
\textbf{* OP-BC Collusion (False Transaction):} In this case, the rogue validators collude to generate and sign a false transaction, verify and validate a block to achieve same BC state. We address this collusion attack using our dynamic validation process which requires validators to verify computational consistency for every successfully validated transaction to reach consensus on the current state of the dynamic block. Unless rogue validators are able to anticipate an accident and send the false transaction before the occurrence of an accident, its of no use sending a false transaction after a primary evidence transaction has been validated in the dynamic block (\textit{dblock}). If however, they collude to periodically send false transaction to be exempted from liability when an accident occurs, a \textit{CAV} owner would be able to detect this since it is authorised to read BC data. \\
\textbf{* DP-BC Collusion (Modification Attack):} in this case, an auto manufacturer colludes with its \textit{CAV} to modify the content of $T_{data}$, computes a new hash value of the data $h_{new}$($T_{data}$) and sends the modified content in a request transaction to the DP-BC. We detect this attack during the verification of transactions sent by all concerned auto manufacturers and insurance companies. Validators comparatively verify the hash of $T_{data}$ presented by all proposers and check the timestamp on the data and location of event. \\
\textbf{Sensor Alteration:} A rogue OP-BC validator could compromise an evidence generating sensor to produce authenticated messages that contain misleading information. To detect this attack, we solely rely on the verification of data consistency. B-FICA comparatively verifies data generated by a \textit{CAV} with data generated by other \textit{CAV}’s involved in the accident contained in the request transaction submitted by DP-BC proposers. Data consistency is also comparatively verified in both time and space by checking data timestamps and location. However, in the case where there are no witnesses, this attack can still go undetected. 
\subsection{Comparative Evaluation}
In this section, we discuss further aspects of B-FICA and comparatively evaluate it with B4F[24] a closely related adjudication framework. The comparison highlights the strengths of our proposed framework which are summarised in Table 2. 
\subsubsection{Proof of storage}
Recall that (Section 3-D), memory intensive data such as pictures and videos are stored in a cloud storage. This data is very useful in B-FICA for making adjudication decisions and processing insurance claims. In order to prove that data stored in the cloud storage has not been altered by the data owner (\textit{CAV} owner) who could be motivated to evade liability, B-FICA stores the hash of the data on the BC for integrity checks and non-repudiation. \\
The authors in [30], [32] conducted an experiment on the cost of storing accident videos in the cloud which is the cost of signing, transmitting, verifying and decrypting the file. In their experiment, they varied the size of accident files between 2 - 8 GB and utilized a data transfer rate of 1.2 GB/s even though it is predicted that the connection speed for 5G-enabled vehicles will be up to 1 Tb/s already achieved for stationary wireless connection [33]. Based on their results, the time overhead to deliver a 2 GB video file requires between 20 seconds to 1 minute and an 8 GB video file requires 1-3 minutes.
\subsubsection{Proof of interaction}
The attribution of liability and processing insurance claims in the \textit{CAV}-setting requires that an adjudicator is able to establish the actual liable entity in case of an accident to prevent wrong decisions. To achieve this, the authors in [9] defined product and negligence liability to identify when a validator is at fault and when a \textit{CAV} owner is at fault. In B-FICA, beyond storing accident-related data in the BC, we also prove that an interaction between liable entities have occurred. In the following, we outline a use case justifying this design choice. Consider a scenario where an auto manufacturer sends an update instruction to a \textit{CAV} owner and the \textit{CAV} owner fails to execute the update instruction and an accident occurs due to that faulty sensor. In alternative proposals such as [24], storing the hash of the update  instruction in the BC serves as proof that the auto manufacturer sent the instruction. However, there is no proof to show that the target \textit{CAV} received and executed the update instruction. This could lead to contestable liability decisions and wrong liability attribution. To address this, B-FICA incorporates an acknowledgement (\textit{ACK}) mechanism which allows a \textit{CAV} owner to confirm receipt of the update instruction sent by the auto manufacturer and a \textit{proof of execution mechanism} which informs BC validators if the transaction was successfully executed or not. This is so that in the use case described above, if a software bug still exists after executing the update, the auto manufacturer would be liable and if no proof of execution is received, the \textit{CAV} owner would be responsible for the accident. 
\subsubsection{Practical benefits of B-FICA}
We illustrate the practical benefits of B-FICA over B4F [24] using the \textit{CAV} accident scenario presented in [1] where the driver was killed in the accident. In this scenario, making adjudication decisions was difficult because evidence was made inaccessible by a competing party. In the existing liability attribution and claims processing model this exceptional scenario could hinder making claims decision for at most 12-months [29]. A variant of this could be that competing parties selectively decide on what data contributes to evidence. B-FICA and B4F[24] both integrate all potential liable entities in the BC network and show comparable performance. However, B4F [24] is both vulnerable to contestable liability decisions as described above and selective data disclosure. Using B4F, a potential liable validator could decide to store only BC transactions that shield it from liability. This is because the validators in B4F separately store their interactions with the vehicle and possess significant control over what data is stored in the BC. An example could be that a service technician after a maintenance activity on a \textit{CAV} includes services it did not render in the maintenance report and stores the hash of the report in the BC. B-FICA on the other hand allows all validators to process all transactions contributing to evidence and using an \textit{ACK} mechanism, transaction participants are apprised on what data is stored in the BC. 
\begin{table*}[ht]
\caption {Comparing B-FICA with B4F [24]} \label{tab:title} 
\centering
\begin{tabular}{|p{2.4cm}|p{1.0cm}|p{0.8cm}|p{12.8cm}|}
\hline
Features & B-FICA & B4F[24]  & Implication and Approach \\
\hline
Liability type (Product and Negligence) [9] & Both & Product & B-FICA presents a holistic approach in the adjudication process. This is because it also keeps track of the interactions between potential liable entities to identify cases of liability due to negligence. It uses multiple signatures (\textit{MultiSig}) to identify negligence and prevent repudiation of actions by ensuring that transaction participants acknowledge received transactions. This is not possible in B4F. \\ 
\hline
Independent storage of evidence & Not possible & Possible & This leads to selective data sharing which allows a validator to influence the outcome of a liability decision. This is akin to the B4F framework where a validator stores its interaction with a \textit{CAV} separately and is expected to store the hash of such transactions in the BC. However in B-FICA all data contributing to evidence is sent to the BC and processed by all validators. This is because while B4F treats validators as trusted entities, we argue that they could be motivated to repudiate their actions. \\ 
\hline
\end{tabular}
\end{table*}
\subsection{Performance Evaluation}
In this section, we evaluate the performance of B-FICA via extensive simulations. We distinctly evaluate the performance of the operational and decision partitions based on transactions processed in the partitions since they independently process transactions. To evaluate B-FICA, we use the NS-3 [16] simulator as it has been extensively utilized as an evaluation tool to analyze peer-to-peer networks. \\
We consider the primary evidence transaction and notification evidence transaction as primary evaluation transactions on which other transactions are predicated. More precisely, the request transaction is initiated after a primary transaction has been received by the OP-BC validators and the execution transaction sent by a \textit{CAV} is based on the notification transaction initiated by an auto manufacturer or service technician. \\
To realistically simulate the primary evaluation transactions, we utilize the road casualty crash data of New South Wales (NSW) [17] to simulate the generation of primary evidence transaction. The crash data report presents the traffic casualties for 2016. It gives a description of the type of crash which describes vehicles involved in a crash and categorizes a crash based on degrees of impact. For notification evidence, we assume weekly updates for connected and automated vehicle. \\
For evaluations, we group the multiple entities in OP-BC into clusters to reflect operational validators with same \textit{CAVs} in a region such as NSW as described in Section 2. For the decision partition, we consider one legal authority and government transport authority for a region. Using the crash casualty data, 42 car accidents occur daily therefore, \textit{CAVs} generate 42 \textit{PET per day} and decision partition validators receive 42 request transactions from insurance companies daily while operational partitions collectively receive 42 transactions. \\
$B_{Max}$ involves a tradeoff between overhead and security. This is because the value of $B_{Max}$  determines when \textit{dBlock} is appended to the BC and becomes immutable. While lower $B_{Max}$ values present larger BC overhead, larger $B_{Max}$ values increase the vulnerability of \textit{dBlock} by offering colluding validators the opportunity to exploit the validation process described in Section 2. Here, we set $B_{Max}$  to 7 and leave the  definition of an optimal $B_{Max}$ value for future work. We assume that at least one \textit{ PET} is generated every hour and a \textit{CAV} receives one notification transaction weekly. We refer to the above experimental settings as default configurations and are used in simulations unless stated otherwise. \\ 
In the rest of the section, we evaluate the time taken to verify and process OP-BC transactions in the dynamic block. In the decision partition, we evaluate the time it takes to process transactions in a block and the time overhead in providing a response to a request evidence. Furthermore, we comparatively evaluate our implementations against a baseline implementation (Section IV-2) and existing proposals. 
\subsubsection{OP-BC Partition Performance Evaluation}
In the operational partition, we evaluate the key processes identified as verification and validation. These processes are critical in  B-FICA because they contribute to how long a complementary evidence is received and determines when a claims decision can be reached.\\
To provide comprehensive simulation, we simulate the OP-BC transactions: primary evidence, notification evidence and execution transactions using the following metrics: \\
\textbf{Verification time: }This refers to the time it takes an OP-BC manager to confirm conformance to verification policies (described in Section 2) after a transaction is received in the BC. \\
\textbf{Validation time: } This refers to the time an OP-BC manager processes a transaction in the dynamic block (\textit{dblock}). It is measured as the time required to compute a new \textit{dblock} ID. \\ 

\begin{figure}[h]
\centering
\includegraphics[width=0.35\textwidth]{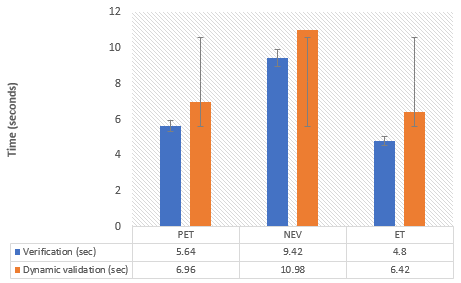}
\caption{Average verification and validation time on OP-BC validators.}
\end{figure}
The results shown in Figure 2 are the average of 14 runs. the standard deviation is also shown. As shown in Figure 2, the notification evidence (\textit{NEV}) has higher processing time compared to other transactions. This is because the \textit{NEV} transaction is a 2-signature transaction which requires additional time for verification checks. Furthermore, most transactions take several seconds to verify and validate. This can be attributed to the storage of transaction data on the chain. First, validators confirm that a transaction meets verification policies explained in Section 2. Second, validators verify the integrity of transaction data to ensure that the data has not been altered during transit by comparing newly computed hash of data to hash of data in the transaction. Finally, after the transaction validation process, validators verify their computations for consistency before a transaction is considered successfully verified and validated. Verification of computational consistency concludes the consensus process and ensures that validators have same computation for \textit{dblock} \textit{ID}. This final step also adds to the overall validation time.  \\
We also compare the verification time to an alternative implementation that stores only hash of transactions in the OP-BC. Results shows that B-FICA introduces additional verification overhead (0.30 seconds for \textit{PET}) which also impacts on how long before an insurance company receives complimentary evidence and can be attributed to the verification of components of transaction data for completeness i.e. for example, checking that $T_{data}$ in \textit{PET} is complete.
\subsubsection{DP-BC Partition Performance Evaluation}
We consider the time overhead in providing response to a request and the time it takes DP-BC validators to process transactions in a block when transactions in the running pool reaches $B_{Max}$. \\
\textbf{Time Overhead: } Time overhead refers to the processing time of the request transaction. It is measured as the time from when a transaction(s) sent from a proposer is received by validators until when the proposer receives an appropriate feedback. \\
\textbf{Block processing time: } This is measured as the time from when a DP-BC validator generates a block, processes transactions in the block and appends the block to the ledger. \\ 
\textbf{Time overhead evaluation: } To evaluate the time overhead, we compare B-FICA against 2 alternative implementations. In the first case called \textit{baseline}, we use similar transaction flow as B-FICA however, we do not hash or encrypt transactions. In the second case, we implemented the idea presented in [24] where actual transaction data are stored in a personal storage and hash of transactions are stored in the BC. \\
\begin{figure}[h]
\centering
\includegraphics[width=0.35\textwidth]{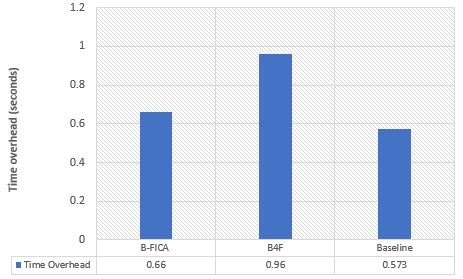}
\caption{Comparative time overhead evaluation of B-FICA.}
\end{figure}
The results in Figure 3 shows that B4F expends the most time in processing transactions and providing feedback to the transaction proposer. This increased overhead can be because of the additional integrity verification checks conducted by DP-BC validators on data retrieved from the personal storage before providing response to the transaction proposer. Integrity verification hash involves computing the hash of the transaction in the personal storage and comparing it with the hash of transaction in the BC. This computation time therefore adds to the time overhead. Furthermore, compared to the baseline, B-FICA includes an additional overhead of 0.13 seconds to process transactions. This overhead which could be called security cost is the time taken for the encryption and hashing operations. \\
\textbf{Block processing time: } It takes a DP-BC B-FICA validator 56.2 seconds to process transactions in $B_{Max}$. Compared to  the alternative hash only implementation, it takes about similar time (56.3 seconds). This is because similar computation is performed to ensure DP-BC validators reach consensus (see Section 2) on the state of the DP-BC. 
\section{Related works}
In this section, we provide a brief review about existing works on vehicular forensics, BC and its application in the vehicular networks domain.  
\subsection{Vehicular Forensics}
Existing vehicular technologies enables a range of applications through vehicular ad-hoc networks (VANETs). Specifically, applications needed for vehicles to confirm their current state or prove their recent behaviors. For example, when vehicle collision occurs, it is important to maintain an accurate description of the situation for liability purposes. \\
The authors in [18] proposed a cooperative collision warning based motor vehicle event data recorder (MVEDR) for recording all critical information from a vehicle in an accident (primary vehicle) and neighbor vehicles for accident reconstruction. However, while MVEDR represents a vital solution for accident reconstruction for modern day vehicular forensics, it solely relies on data in the EDR which records only the perception of a host vehicle. This therefore makes it unsuitable for the anticipated \textit{CAV} adjudication model where liability is likely to be shared between multiple entities. \\
The author in [19] proposed EVIGEN protocol for obtaining data about a vehicles behavior from nearby vehicles. It involves creating a digital document to attest to the data collected for adjudication. While EVIGEN represents an overlay of MVEDR, it however relies on the availability of witnesses to obtain reliable data for adjudication.\\
The authors in [3], [20] have also  proposed forensic VANET applications to facilitate accident reconstruction. They identified the limitations of alternative proposals and defined dual sources of evidence for reliable data collection for accident reconstruction. In their proposal, a host vehicle can record its perception of an accident and also store data from neighboring witnesses. However, the challenge is that evidence retrieved is stored in different centralized databases which presents a single point of failure [20, 21]. 
\subsection{Blockchain Applications for Vehicular Networks}
In the vehicular networks domain, BC has been proposed for different purposes. The author in [22] proposed Blackchain, a BC based message and revocation accountability system for secure vehicular communication. The authors in [23] presents a BC based secure architecture for automotive use cases such as dynamic vehicle insurance which allows an insurance company to provide specific services to drivers by monitoring their driving patterns. However their proposal lacks a concrete example on how this will be achieved. The authors in [24] proposed Block4Forensic, a BC based solution for vehicular forensics that integrates concerned entities and facilitates the provision of necessary data needed for dispute settlement. However, it does not cover all liability types as specified in [26] for self driving vehicles therefore, we consider our proposal a holistic solution for vehicular forensics. 
\section{Conclusion}
In this paper, we proposed a Blockchain based Framework for auto-Insurance Claims and Adjudication (B-FICA) for connected and automated vehicles (\textit{CAVs}). B-FICA enforces access control using partitions to prevent unauthorised access to data contributing to evidence and uses a dynamic validation protocol to prevent evidence alteration. Using a likely scenario, we demonstrated the efficacy of B-FICA describing how evidence is generated and processed. Security analysis demonstrates mechanisms utilised by B-FICA to achieve design requirements and its resilience against a broad range of malicious actions of potential liable entities. Simulation results shows that the proposed framework significantly decreases processing time compared to the state of the art. Generally, B-FICA brings credibility to the adjudication model for \textit{CAVs} and prevents data misuse at a marginal cost. \\
Our current work provides comprehensive information to decision makers, proves interaction between entities in the adjudication model and prevents potential liable entities from altering submitted evidence. An interesting direction for future work is to implement attestation protocols to ensure the correctness, reliability and integrity of data submitted for evidence by a connected and automated vehicle. Another future direction is to relax the assumption that the auto manufacturer provides the software needs of a \textit{CAV} and consider third party providers. We expect variations of B-FICA to be applicable in other liability scenarios, such as home or health insurance.
\section{Acknowledgement}
The authors thank the editors and the anonymous reviewers for their comments and insightful feedbacks. This work is supported by Commonwealth Scientific and Industrial Research Organization's (CSIRO) data61 and the University of New South Wales. 

\begin{thebibliography}{1}
\bibitem{IEEEhowto:kopka}
N.~Boudette, \emph{Autopilot cited in Death of Chinese Tesla Driver}, \relax Also available at:
https://www.nytimes.com/2016/09/15/business/fatal-tesla-crash-in-china-involved-autopilot-government-tv-says.html \

\bibitem{IEEEhowto:kopka}
Gabler, Hampton C, C. Hampton and T.A Roston, \emph{Estimating Crash Severity: Can Event Data Recorders Replace Crash Reconstruction?} \relax (Proceedings of 18th International Technical Conference of Safety of Vehicles,  2003. \

\bibitem{IEEEhowto:kopka}
Y. Kopylova, C. Farkas, and W. Xu,   \emph{Accurate Accident Reconstruction in VANET.} \relax Data and Applications Security and Privacy XXV. Springer Berlin Heidelberg, 2011. 271-279. \

\bibitem{IEEEhowto:kopka}
Richard Boon, \emph{Post-accident Analysis of Digital Sources for Traffic Accidents.} \relax 21st Twente Student Conference on IT, June 23rd, 2014, Enschede, The Netherlands. \

\bibitem{IEEEhowto:kopka}
S. Nakamoto, \emph{Bitcoin: A peer-to-peer electronic cash system.} \relax 2008. \

\bibitem{IEEEhowto:kopka}
Hyperledger Research Community,\emph{Welcome to Hyperledger Fabric.} \relax Available on: https://hyperledger-fabric.readthedocs.io/en/latest/  \ 

\bibitem{IEEEhowto:kopka}
C.~Oham, S. S.~Kanhere, R.~Jurdak and S.~Jha, \emph{A Blockchain Based Liability Attribution Framework for Autonomous Vehicles.} \relax  Also available on: https://arxiv.org/abs/1802.05050   \ 

\bibitem{IEEEhowto:kopka}
T.~Simonite, \emph{Tesla Knows When a Crash Is Your Fault, and Other Carmakers Soon Will, Too.} \relax June, 2016. \

\bibitem{IEEEhowto:kopka}
Norton Rose Fulbright, \emph{Autonomous vehicles: The legal landscape of dedicated short range communication in the US, UK and Germany.} \relax July, 2017.  \ 

\bibitem{IEEEhowto:kopka}
ETSI, \emph{Intelligent Transport System (ITS), Vehicular Communications; Basic Set of Applications; Part 3: Specifications of Decentralized Environmental Notification Basic Service.} \relax ETSI TS, 102 637-3 V1.1.1 (2010-09). \

\bibitem{IEEEhowto:kopka}
NRMA Insurance,\emph{Car Insurance Claims.} \relax Also available on: https://www.nrma.com.au/claims/car-insurance   \ 

\bibitem{IEEEhowto:kopka}
DMV.org,  \emph{How to handle staged car accidents} \relax Available on:https://www.dmv.org/insurance/how-to-handle-staged-car-accidents.php  \ 

\bibitem{IEEEhowto:kopka}
B.~Sage,  \emph{Insurance Scam Fraud Protection} \relax Available on: https://www.thebalance.com/insurance-scam-fraud-protection-2645469  September, 2016. \ 

\bibitem{IEEEhowto:kopka}
C.~ Hammerschmidt, \emph{The best way to attack PoW blockchains for profit.} \relax Also available on: https://medium.com/@chrshmmmr/a-guide-to-dishonesty-on-pow-blockchains-when-does-double-spending-pays-off-4f1994074b52 March, 2017. \

\bibitem{IEEEhowto:kopka}
Investopedia,  \emph{Reputational Risk.} \relax Also available on: https://www.investopedia.com/terms/r/reputational-risk.asp   \ 

\bibitem{IEEEhowto:kopka}
NS3,  \emph{Documentation.} \relax Also available on: https://www.nsnam.org/ns-3-dev/documentation/   \ 

\bibitem{IEEEhowto:kopka}
NSW Government,  \emph{Road Traffic Casualty Crashes in New South Wales: Statistical Statement for the Year Ended 31st December, 2016.} \relax Transport for NSW, center for road safety, Page 47.    \ 

\bibitem{IEEEhowto:kopka}
C. P.~Young, B. R.~Chang, J.J.~Lin and R. Y.~Fang,\emph{Cooperative Collision Warning Based Highway Vehicle Accident Reconstruction.} \relax Eight International Conference on Intelligent Systems Design and Applications. 2008.  \ 

\bibitem{IEEEhowto:kopka}
J. Fuentes, A. Tablas, A. Ribagorda,   \emph{Witness-based Evidence Generation in Vehicular Ad-hoc Networks.} \relax ESCAR, 2009.\

\bibitem{IEEEhowto:kopka}
S. Rahman and U. Hengartner, \emph{Secure Crash Reporting in Vehicular Ad-hoc Networks.} \relax University of Waterloo, 2017.\

\bibitem{IEEEhowto:kopka}
Z. Liu, J. Ma, Z. Jiang, H. Zhu and Y. Miao, \emph{LSOT: A Lightweight Self-Organized Trust Model in VANETs.} \relax Mobile Information Systems, 2016.\

\bibitem{IEEEhowto:kopka}
R. Heijden, F. Engelmann, D. Modinger, F. Schonig and F. Kargl, \emph{Blackchain: Scalability for Resource-Constrained Accountable Vehicle-to-X Communication.} \relax SERIAL’17: ScalablE and Resilient InfrAstructures for distributed Ledgers, December 11–15, 2017. \

\bibitem{IEEEhowto:kopka}
A. Dorri, M. Steger, S. Kanhere and R. Jurdak, \emph{BlockChain: A Distributed Solution to Automotive Security and Privacy.} \relax IEEE Communication Magazine, 2017. \

\bibitem{IEEEhowto:kopka}
M. Cebe, E. Erdin, K. Akkaya, H. Aksu and S. Uluagac, \emph{Block4Forensic: An Integrated Lightweight Blockchain Framework for Forensics Applications of Connected Vehicles.} \relax  Also available at: https://arxiv.org/pdf/1802.00561.pdf  2018. \

\bibitem{IEEEhowto:kopka}
A.~Kulkarni, \emph{How To Choose Between Public And Permissioned Blockchain For Your Project.} \relax Also available on: https://blog.chronicled.com/how-to-choose-between-public-and-permissioned-blockchain-for-your-project-3c5d4796e3c8  \ 

\bibitem{IEEEhowto:kopka}
Norton Rose Fulbright, \emph{Autonomous vehicles: The legal landscape of dedicated short range communication in the US, UK and Germany.} \relax July, 2017.  \ 
\bibitem{IEEEhowto:kopka}
IBM Research, Zurich \emph{Specification of the Identity Mixer Cryptographic Library.} \relax April, 2010.  \ 

\bibitem{IEEEhowto:kopka}
R.~Tummala,  \emph{Autonomous Cars: Radar, Lidar, Stereo Cameras.} \relax Georgia Tech Packaging Research Center, IEEE-CPMT Workshop - Autonomous Cars. \ 

\bibitem{IEEEhowto:kopka}
Financial Rights Legal Centre, \emph{Help! My insurance claim is taking forever!} \relax Insurance Law Service - Advice and advocacy for consumers in financial stress. 2017.\


\bibitem{IEEEhowto:kopka}
M.~Eiza, Q.~Ni and Q.~Shi  \emph{Secure and Privacy-Aware Cloud Assisted Video Reporting Service in 5G Enabled Vehicular Networks.} \relax IEEE Transactions on Vehicular Technology VTSI-2015-01271.R1. 2015 \ 

\bibitem{IEEEhowto:kopka}
P.~Kamat, A.~Baliga and W.~Trappe  \emph{Secure, pseudonyms, and auditable communication in vehicular ad hoc networks.} \relax Security and Communication Networks, 2008. \ 

\bibitem{IEEEhowto:kopka}
L.~Nkenyereye, J.~Kwon and Y.~Choi \emph{Secure and Lightweight Cloud-Assisted Video Reporting Protocol over 5G-Enabled Vehicular Networks.} \relax Sensors, MDPI 2017. \ 

\bibitem{IEEEhowto:kopka}
BBC News, \emph{5G researchers manage record connection speed - BBC News.} \relax Also available online: http://www.bbc.co.uk/news/technology-31622297. 2015 \ 
\end{thebibliography}
\end{document}